\pdfoutput=1
\documentclass[11pt]{article}

\usepackage[utf8]{inputenc}
\usepackage[T1]{fontenc}
\usepackage{amsmath,amssymb}
\usepackage{graphicx}
\usepackage{booktabs}
\usepackage[colorlinks=true,
linkcolor=blue,
citecolor=blue,
urlcolor=blue]{hyperref}
\newcommand{\doi}[1]{\href{https://doi.org/#1}{doi:#1}}
\usepackage[numbers,sort&compress]{natbib}       
\usepackage{authblk}
\usepackage{geometry}
\title{An Adaptive, Data-Integrated Agent-Based Modeling Framework for Explainable and Contestable Policy Design}

\author{
	Roberto Garrone$^{1,2}$\thanks{ORCID: \href{https://orcid.org/0009-0005-7060-6774}{0009-0005-7060-6774}}\\
	$^{1}$Department of Informatics, Systems and Communication (DISCo), University of Milano--Bicocca, Milan, Italy\\
	$^{2}$Faculty of Pure and Applied Sciences, Open University of Cyprus, Nicosia, Cyprus
}

\date{November 24, 2025}

\begin{document}
	
	\maketitle
	
	\begin{abstract}
		Multi-agent systems often operate under feedback, adaptation, and non-stationarity, yet many simulation studies retain static decision rules and fixed control parameters. This paper introduces a general adaptive multi-agent learning framework that integrates: (i) four dynamic regimes distinguishing static versus adaptive agents and fixed versus adaptive system parameters; (ii) information-theoretic diagnostics—entropy rate, statistical complexity, and predictive information—to assess predictability and structure; (iii) structural causal models for explicit intervention semantics; (iv) procedures for generating agent-level priors from aggregate or sample data; and (v) unsupervised methods for identifying emergent behavioral regimes. The framework offers a domain-neutral architecture for analyzing how learning agents and adaptive controls jointly shape system trajectories, enabling systematic comparison of stability, performance, and interpretability across non-equilibrium, oscillatory, or drifting dynamics. Mathematical definitions, computational operators, and an experimental design template are provided, yielding a structured methodology for developing explainable and contestable multi-agent decision processes.
	\end{abstract}
	

\section{Introduction}

Adaptive multi-agent systems (MAS) increasingly underpin decision processes in
domains such as energy, mobility, environmental regulation, and public policy.
Agents interact, adapt, and respond to evolving system parameters, while
policymakers revise controls in response to observed performance. These
socio-technical systems exhibit feedback, path dependence, and emergent
structure. Yet methodological tools for jointly studying agent adaptation,
policy learning, and system-level dynamics remain fragmented across behavioral
modeling, reinforcement learning, causal inference, and complex systems
analysis.

This paper proposes a unified framework for analyzing adaptive MAS that
integrates four methodological pillars. First, a regime-based architecture
distinguishes between static and adaptive agents and between fixed and adaptive
policy parameters. The resulting four regimes---CPCA, CPVA, VPCA, and VPVA---form
a conceptual map for comparing MAS configurations with heterogeneous degrees of
adaptation and feedback.

Second, the framework incorporates a transparent behavioral layer that allows
agents to form and revise \emph{beliefs about policy trajectories}. Belief-driven
adaptation provides an interpretable alternative to opaque learning rules: agents
react not only to instantaneous policy values but also to perceived patterns in
policy evolution. This preserves bounded rationality while enabling structured
reactivity, and it supports causal and counterfactual analysis of agent responses.

Third, we introduce a declarative specification layer for representing policy
rules, causal pathways, and intervention semantics. Using a lightweight,
rule-based formalism, policymakers can articulate constraints, goals, and causal
assumptions. Agents may access a restricted, policy-only subset of this
representation, bridging symbolic and numerical perspectives and enhancing
contestability and transparency.

Fourth, we integrate diagnostic tools from information theory, causal inference,
and unsupervised learning. Entropy rate, statistical complexity, and predictive
information quantify the structure and predictability of emergent trajectories.
Structural causal models (SCMs) provide explicit semantics for interventions and
counterfactual reasoning. Clustering methods identify distinct behavioral or
policy regimes arising from the interaction of adaptation and control.

Together, these components form a general, domain-neutral architecture for
studying adaptive MAS without presupposing convergence or equilibrium. The
framework supports systematic comparison across static, semi-adaptive, and fully
adaptive configurations, enabling researchers and policymakers to evaluate
stability, interpretability, and robustness. By integrating belief formation,
declarative causal specification, and information-theoretic diagnostics, the
framework contributes to the foundations of explainable and contestable
multi-agent decision systems.

The contributions of this work are:

1. A four-regime architecture for adaptive multi-agent systems integrating agent
learning and adaptive policy search;
2. A belief-driven behavioral layer capturing interpretable agent reactions to
evolving policies;
3. A declarative, rule-based specification layer for causal pathways and
intervention semantics;
4. A unified diagnostic suite combining structural causal models and
information-theoretic measures;
5. Three policy-relevant instantiations (load balancing, smart grids, emissions).

	\section{Background and Related Work}
	
	\subsection{Multi-Agent Systems and Agent-Based Models}
	
	
	Multi-agent systems provide a general paradigm for modeling distributed
	decision-making, where multiple interacting entities pursue goals under
	partial information, limited rationality, and feedback. Classical work in
	distributed AI and coordination established foundational principles of
	interaction, cooperation, and communication in MAS
	\citep{durfee1988,shoham1997,wooldridge2009}.
	Within AI, MAS research spans game-theoretic interaction
	\citep{osborne1994,myerson1997}, cooperative and competitive multi-agent
	reinforcement learning \citep{busoniu2008,zhang2021}, and decentralized
	control and planning under uncertainty
	\citep{oliha2000,bernstein2002,oliehoek2016}.
	These traditions emphasize how local decision rules, information
	constraints, and coordination mechanisms shape emergent global behavior
	in distributed systems.
	
	Agent-based modeling (ABM) is closely linked to this tradition and to the
	study of complex adaptive systems (CAS), where macro-level order emerges
	from micro-level interactions under adaptation and feedback
	\citep{holland1992,miller2007,levin1998}. In generative social science
	\citep{epstein1999,epstein2012,epsteinaxtell1996}, explanation is achieved
	by constructing mechanisms that reproduce empirical patterns, and ABMs have
	been widely used in computational social science
	\citep{cioffi2017,conte2014,gilbert2008} to represent heterogeneous agents,
	bounded rationality, and networked interaction.
	
	Applications span diverse domains (e.g., epidemiology, mobility, markets,
	and public services), where ABMs and MAS are used as “laboratories” to
	explore consequences of alternative designs or interventions
	\citep{dunham2005,germann2006,epstein2009,perez2009,ajelli2010,
		auchincloss2019,comis2021,tracy2018,gostoli2019,colombo2011,
		shergold2012,boyd2022}. These examples motivate the need for general
	methodologies that can handle adaptation, non-stationarity, and feedback
	without being tied to any single application.
	
	\subsection{Synthetic Populations and Agent Initialization}
	
	Realistic MAS and ABMs often require plausible agent populations. Synthetic
	population methods, beginning with \citet{beckman1996}, use iterative
	proportional fitting (IPF) to reweight sample microdata so that marginal
	distributions match aggregate constraints. Recent surveys review extensions
	and good practices \citep{chapuis2022,predhumeau2023,vonhoene2025}. Multiple
	imputation \citep{rubin1996} addresses missing attributes and facilitates
	uncertainty analysis.
	
	In an AI context, these techniques can be viewed as generic procedures for
	constructing heterogeneous agent priors: given aggregate constraints and a
	sample (or proxy) dataset, they produce a distribution over agent-level
	attributes that can feed any downstream learning or decision process.
	This perspective abstracts away from specific domains and treats synthetic
	populations as a modular component of agent initialization.
	
	\subsection{Structured Interaction Topologies}
	
	Interaction structures in MAS are naturally represented as graphs or
	networks. GIS-informed ABMs and spatial MAS embed these networks in
	geographic space \citep{crooks2008,heppenstall2016,banos2015}, but more
	generally one may consider abstract interaction topologies $G=(V,E)$ with
	attributes on nodes and edges. Such structures govern who can interact with
	whom, what information flows where, and how costs or constraints (distance,
	capacity, congestion) shape behavior.
	
	Networked interaction is central to many AI applications: distributed
	sensing, communication networks, multi-robot systems, and social or
	information networks. The present framework treats interaction topology as
	a first-class object, independent of any specific spatial embedding.
	
	\subsection{Validation, Sensitivity Analysis, and Documentation}
	
	Structured validation practices have been proposed to improve the
	credibility of ABMs and MAS-based simulations. Recent work emphasizes
	conceptual, empirical, and predictive validity, as well as best practices
	for reporting \citep{collins2024,an2021}. Sensitivity analysis techniques,
	including Morris screening \citep{morris1991} and variance-based methods
	such as Sobol indices \citep{saltelli2010}, help identify influential
	parameters and quantify uncertainty. Classical simulation studies also
	address initialization bias and steady-state analysis
	\citep{schruben1983,kleijnen1984,kleijnen2007}. The ODD protocol provides
	standardized documentation to facilitate replication and transparency
	\citep{grimm2020}.
	
	These ideas carry over directly to AI-driven MAS: learning rules and
	control parameters can be subjected to the same systematic experimental
	design, and validation can be framed in terms of predictive performance,
	structural robustness, and invariances across interventions.
	
	\subsection{Causality, Information Theory, and Explainable Multi-Agent Systems}
	
Traditional ABMs explain outcomes mechanistically but rarely encode explicit causal structures. In AI, there is a growing interest in combining structural causal models (SCMs) with learning systems to clarify what is being assumed and what is being learned \citep{pearl2009,hernan2020,rothman2008,vdw2015,hill1965,rubin1974}. Recent contributions integrate SCMs and intervention logic with simulation models \citep{chang2024,schluter2023,rieder2025}, aligning MAS with modern causal inference. Contestability—the capacity for stakeholders to scrutinize, challenge, and understand model assumptions and outputs—is recognised as a core requirement for accountable systems \cite{selbst2019fairness,euaiact2024}. Incorporating causal graphs, explicit assumptions, and diagnostic metrics helps operationalise this requirement.

Computational mechanics and information theory provide tools to quantify emergent structure and predictability \citep{crutchfield1989,crutchfield1994,shalizi2001,cover2006,li2008}. Metrics such as entropy rate, statistical complexity, and predictive information can distinguish between randomness, simple deterministic dynamics, and complex structure. When applied to MAS, they allow one to characterize different learning and control regimes in terms of information storage and predictability, offering a basis for explainability and model comparison.

From a design perspective, adaptive control and resilience have become central concerns in AI-supported decision systems. MAS are increasingly used as testbeds where interventions and learning strategies can be evaluated under controlled conditions and revised iteratively \citep{douglas2019,keck2013,northridge2016}.
	
	\section{A General Adaptive Multi-Agent Framework}
	
	\subsection{Conceptual Overview}
	
	We consider a generic multi-agent decision system comprising a population
	of agents, an environment, and a set of system-level control parameters.
	The proposed framework is structured into five layers:
	
	\begin{enumerate}
		\item \textbf{Population layer}: synthetic agents generated via IPF and imputation, informed by surveys or sample data.
		\item \textbf{Environment layer}: spatial or abstract network topology constraining interactions.
		\item \textbf{Behavioral layer}: agent decision rules, static or adaptive.
		\item \textbf{Control layer}: a vector of system-level parameters, static or subject to search.
		\item \textbf{Diagnostics layer}: performance metrics, causal graphs, information-theoretic
		measures, and emergent pattern analysis via clustering.
	\end{enumerate}
	
	Within this structure, we distinguish four dynamic regimes that define how
	agents and control parameters co-evolve.
	
	\subsection{Four Dynamic Regimes}
	
	Let $s_t \in S$ denote the system state at discrete time $t$, $A = \{a_i\}$
	the set of agents, and $P_t \in \mathbb{R}^d$ a vector of control
	parameters. Each agent $i$ has an internal state $\theta_{i,t}$ and chooses
	an action $x_{i,t} \in X_i$ according to a behavioral rule $R_i$. In the
	adaptive case, internal states update according to a learning rule $L_i$.
	
	Agents act in an environment defined by a spatial or network topology (see
	Section \ref{sec:spatial}). The transition function $F$ maps current state,
	actions, and control to the next state:
	\[
	s_{t+1} = F(s_t, X_t, P_t, \zeta_t),
	\]
	where $\zeta_t$ captures exogenous shocks.
	
	Control parameters may be fixed or updated via an optimization rule $G$:
	\[
	P_{t+1} = G(P_t, \hat{J}_t, s_t),
	\]
	where $\hat{J}_t$ is an intermediate performance estimate.
	
	Combining static vs.\ adaptive agents and fixed vs.\ adaptive control yields
	four regimes:
	
	\begin{itemize}
		\item \textbf{CPCA (Constant Policy, Constant Agents)}: $P_t \equiv P$, $L_i = \varnothing$.
		\item \textbf{CPVA (Constant Policy, Variable Agents)}: $P_t \equiv P$, $L_i \neq \varnothing$.
		\item \textbf{VPCA (Variable Policy, Constant Agents)}: $L_i = \varnothing$, $P_t$ updated by $G$.
		\item \textbf{VPVA (Variable Policy, Variable Agents)}: both $L_i \neq \varnothing$ and $P_t$ updated.
	\end{itemize}
	
	The framework treats all four regimes within a unified notation, allowing
	systematic comparison of stability and performance properties across
	different combinations of agent learning and system-level adaptation.
	
	\subsection{Performance Evaluation Under Non-Convergent Dynamics}
	\label{sec:performance}
	
	Let $\Phi(s_t)$ be a bounded performance functional (e.g.\ combining
	efficiency, equity, and stability objectives). Over a window of length $K$,
	the performance of a control--learning configuration $(P,L)$ is
	\[
	J(P;L) = \frac{1}{K} \sum_{t=T-K+1}^T \Phi(s_t),
	\]
	for a finite simulation horizon $T$. This definition does not require
	convergence to a fixed point; it remains well-defined under stationary,
	cyclic, or drifting dynamics, provided state variables are bounded. Multiple
	replications with different random seeds yield an empirical distribution of
	$J(P;L)$, from which means and variances can be estimated.
	
	\section{Population Layer: Synthetic Populations and Survey Priors}
	
	\subsection{Synthetic Populations via Iterative Proportional Fitting}
	
	Synthetic populations approximate real-world heterogeneity while preserving
	confidentiality \citep{beckman1996,chapuis2022,predhumeau2023,vonhoene2025}.
	Given aggregate marginals (e.g.\ counts by age, income, or other
	categories) and a sample microdataset, IPF reweights micro records so that
	the resulting synthetic population matches the marginals. Let $\mathbf{w}$
	denote weights over sample records; IPF iteratively adjusts $\mathbf{w}$ to
	match each marginal distribution in turn.
	
	In the proposed framework, IPF is used in a domain-neutral way: the same
	method can be applied to any context where aggregate constraints and
	microdata (or a proxy dataset) are available. Multiple imputation
	\citep{rubin1996} can augment the synthetic population with missing
	attributes and encode uncertainty, yielding an ensemble of plausible agent
	initializations.
	
\subsection{Survey-Informed Behavioral Priors}

Survey data provide empirical distributions for attitudes, preferences, expectations, 
and behavioral dispositions, making them a natural source of priors for initializing 
heterogeneous agents. Foundational behavioral theories demonstrate that survey-measured 
attitudes and intentions are systematically linked to action \citep{ajzen1991}, while 
behavioral game theory shows how risk aversion, reciprocity, compliance tendencies, 
and responsiveness to incentives can be elicited empirically and incorporated into 
decision models \citep{camerer2005}. In agent-based modeling, survey responses have 
long been used to parameterize heterogeneity in thresholds, personality traits, and 
behavioral propensities \citep{balke2014}, providing realistic distributions over 
agent-level parameters.

From a methodological perspective, survey data are widely recognized as a reliable 
means of capturing behavioral constructs and subjective expectations 
\citep{nowak2017}, especially when used to shape priors rather than impose strict 
deterministic rules. These priors inform the initial distribution of internal states 
$\theta_{i,0}$—for example, attitudes toward compliance, risk tolerance, preference 
weights, or technology adoption—and may influence learning rates or thresholds in 
$L_i$, thereby conditioning early-stage dynamics. Generative social science further 
emphasizes that such empirically grounded heterogeneity is essential for producing 
plausible emergent macro-structures \citep{epstein2006}. In this framework, surveys 
are therefore treated in a domain-neutral manner as structured sources of prior 
distributions that shape agent initialization and subsequently interact with the 
learning and adaptation dynamics of multi-agent systems.

	\section{Environment Layer: Spatial and Network Structures}
	\label{sec:spatial}
	
	Spatial and network structures are critical in many multi-agent decision
	systems. In spatial MAS and GIS-informed ABMs, environments are represented
	using nodes (locations) and edges (connections), possibly embedded in
	geographic space \citep{crooks2008,heppenstall2016,banos2015}. More
	generally, the framework uses an abstract representation: an environment is
	a graph $G = (V,E)$, optionally with geometric coordinates and attributes
	on nodes and edges.
	
	Agents occupy or traverse nodes, interact with neighbors, and experience
	costs or constraints (e.g.\ distance, congestion, capacity). This structure
	is applicable to mobility, resource distribution, information flows, and
	many other MAS settings, whether or not they have an explicit spatial
	embedding.
	
	\section{Behavioral and Control Layers: Static vs.\ Adaptive Dynamics}
	
	\subsection{Agent Learning}
	
	In the static case, an agent $i$ follows a fixed rule
	$R_i(x_{i,t}, s_t, P_t)$; in the adaptive case, an internal state
	$\theta_{i,t}$ updates according to a learning rule
	\[
	\theta_{i,t+1} = L_i(\theta_{i,t}, s_t, P_t, x_{i,t}, r_{i,t}),
	\]
	with $r_{i,t}$ a realized payoff. This formulation encompasses boundedly
	rational adaptive rules, simple reinforcement learning schemes
	\citep{arthur1994,charpentier2023}, and other heuristics used in MAS and
	ABM to model learning and adaptation.
	
	\subsection{Control (Policy) Search}
	
	Control or policy search treats the MAS as a noisy black-box mapping
	$P \mapsto J(P;L)$ \citep{epsteinaxtell1996,oremland2014,tesfatsion2006}.
	An external optimizer updates $P_t$ based on performance estimates. A
	simple hill-climbing algorithm explores a neighborhood of $P_t$ and moves
	to candidates with higher $J$ if improvements exceed a tolerance. More
	sophisticated search procedures (e.g.\ evolutionary algorithms, Bayesian
	optimization, policy gradient methods) can be plugged into the same
	architecture.
	
	\subsection{Evaluation and Optimization Algorithms}
	
	A generic evaluation procedure runs $R$ replications of the MAS for a given
	$(P,L)$, computes $J(P;L)$ for each replication, and returns mean and
	variance. An optimization procedure iteratively calls the evaluation
	routine for neighboring control vectors until no further improvement is
	detected. These algorithms are modular and apply to all four regimes,
	allowing the framework to be used both for analysis of fixed designs and
	for explicit control optimization.
	
	\subsection{Belief-Driven Behavioral Adaptation}
	
	To align with explainable and model-driven agent architectures, we extend the
	behavioral layer with a lightweight belief model. Agents do not form beliefs
	about other agents or the full environment; instead, each agent maintains
	simple, bounded beliefs about the policy vector $P_t$.
	
	Let $b_{i,t}(P)$ denote agent $i$’s belief distribution over policy parameters.
	Agents update beliefs using observed policy changes:
	\[
	b_{i,t+1}(P) = H_i\bigl(b_{i,t}(P), P_t, \Delta P_t, s_t\bigr),
	\]
	where $H_i$ is an update rule combining prior beliefs and recent policy moves
	(e.g., a Bayesian update, exponential smoothing, or threshold-triggered
	revisions).
	
	Beliefs influence emissions-, consumption-, or demand-generating actions:
	\[
	x_{i,t}
	= f(\theta_i, \eta_i, b_{i,t}(P), s_t).
	\]
	
	This modification preserves bounded rationality and avoids full-blown strategic
	reasoning while enabling agents to respond to perceived policy trajectories.
	Belief updating also improves interpretability: agents adapt to the \emph{pattern}
	of policies, not only to the instantaneous values of $P_t$, producing dynamics
	amenable to causal and information-theoretic analysis.
	
	\subsection{Declarative Specification of Policies and Causal Pathways}
	
	To increase transparency and contestability, we introduce a declarative view of
	policy and causal assumptions. Let $\mathcal{L}$ be a rule-based language over a
	set of predicates representing policy parameters, agent attributes, causal
	links, and admissible interventions. A declarative policy specification has the form:
	\[
	\text{rule: } \mathrm{policy\_update}(P_{t+1}) \leftarrow 
	\mathrm{state}(s_t),\, \mathrm{goal}(G),\, \mathrm{constraint}(C).
	\]
	
	Causal pathways are encoded as logical clauses:
	\[
	\mathrm{causes}(P_t, E_t) \leftarrow \mathrm{mechanism}(M),\ \mathrm{context}(K),
	\]
	which corresponds to structural equations in the SCM.
	
	Agents may access a restricted, policy-only subset of $\mathcal{L}$,
	denoted $\mathcal{L}_P$. This allows them to form beliefs based on declarative
	statements such as:
	\[
	\mathrm{expected\_increase}(\lambda) \leftarrow \mathrm{trend}(P_{t-3:t}).
	\]
	
	The declarative layer does not replace numerical simulation; rather, it serves
	as an interpretable scaffold for specifying intervention semantics, policy
	transitions, and causal assumptions. It bridges ABM dynamics with symbolic
	explanation models and supports the contestability requirements of policy
	simulation.

	\section{Diagnostics Layer: Causality, Information, and Emergent Patterns}
	
	\subsection{Information-Theoretic Measures}
	
	Time series from simulation outputs can be analyzed using
	information-theoretic measures \citep{crutchfield1989,crutchfield1994,
		shalizi2001,cover2006}:
	
	\begin{itemize}
		\item \textbf{Entropy rate} $h_\mu$: asymptotic unpredictability per time step.
		\item \textbf{Statistical complexity} $C_\mu$: amount of information stored in the causal state
		representation.
		\item \textbf{Predictive information} $E$: mutual information between past and future.
	\end{itemize}
	
	These quantities distinguish between random, ordered, and complex regimes,
	and can reveal when control adjustments or learning rules move the system
	toward more predictable or more chaotic behavior. They provide an information-theoretic lens on multi-agent learning dynamics and recent work operationalizes these diagnostics specifically within adaptive MAS via reconstructed $\epsilon$-machines and Kolmogorov-style state compression \citep{garrone2025}.
	
	\subsection{Structural Causal Models and Counterfactuals}
	
	Structural causal models (SCMs) \citep{pearl2009,hernan2020,rothman2008,
		vdw2015,hill1965,rubin1974} represent variables and interventions via
	directed acyclic graphs and structural equations. In the proposed
	framework, SCMs are used to:
	
	\begin{itemize}
		\item clarify assumed pathways through which control variables affect outcomes;
		\item define do-operator interventions corresponding to changes in control parameters;
		\item support counterfactual queries about alternative choices of system-level parameters.
	\end{itemize}
	
	Micro-level mechanisms in the MAS provide dynamics consistent with the
	SCM, while SCMs supply a transparent, contestable representation of causal
	assumptions. Coupling MAS with SCMs thus supports explainable and
	contestable decision-support systems.
		
	\subsection{Clustering and Emergent Regimes}
	
High-dimensional simulation outputs (e.g.\ distributions of indicators
across agents, locations, and time) are hard to interpret visually.
Unsupervised learning techniques---principal component analysis (PCA)
\citep{jolliffe2016}, t-SNE \citep{maaten2008}, $k$-means clustering
\citep{arthur2007}, and Gaussian mixture models \citep{dempster1977}—
can identify emergent regimes and reduce dimensionality. Applications
in ABM and complex-systems research show that clustering can reveal
qualitatively distinct behavioral patterns \citep{hunter2017,fougeres2012,
	grazzini2017,edmonds2005}, enabling systematic interpretation of model
trajectories. The framework leverages these
	tools to:
	
	\begin{itemize}
		\item group simulation runs into archetypal behaviors (e.g.\ stable vs.\ unstable, concentrated vs.\ dispersed);
		\item connect clusters with parameter configurations and dynamic regimes;
		\item support qualitative interpretation and communication of results.
	\end{itemize}
	
	Together, information-theoretic measures, SCMs, and clustering form a
	diagnostic stack for analyzing MAS trajectories and linking them back to
	learning and control design choices.
	
	\section{Experimental Design}
	
	\subsection{Objectives}
	
	The experimental program is designed to answer the following questions:
	
	\begin{itemize}
		\item How do stability and performance differ across CPCA, CPVA, VPCA, and VPVA?
		\item How do synthetic population heterogeneity, interaction structure, and survey priors affect
		emergent behavior?
		\item How do information-theoretic measures respond to control changes and learning dynamics?
		\item Can clustering reliably identify distinct emergent regimes and relate them to design
		choices?
	\end{itemize}
	
	\subsection{Design and Sampling}
	
	We adopt a computational experimental design. Independent variables include:
	
	\begin{itemize}
		\item regime type (CPCA, CPVA, VPCA, VPVA);
		\item initialization and step sizes of the control vector $P$;
		\item strength and type of learning rules $L_i$;
		\item interaction topology and network density;
		\item degree of heterogeneity in synthetic populations.
	\end{itemize}
	
	For each configuration, multiple replications with different random seeds
	are run for a fixed horizon $T$, and performance is evaluated over a window
	of length $K$ as in Section~\ref{sec:performance}. Parameter sampling may
	use grid or Latin hypercube designs to efficiently cover the space.
	
	\subsection{Analysis Plan}
	
	The analysis will:
	
	\begin{enumerate}
		\item Estimate distributions of $J(P;L)$ by regime and parameter configuration.
		\item Assess stability via classification of trajectories (stationary, cyclic, drifting).
		\item Compute entropy rate, $C_\mu$, and predictive information across runs.
		\item Use Morris and Sobol indices to identify influential parameters.
		\item Apply clustering to aggregate output statistics and identify emergent regimes.
		\item Map clusters back to control and learning configurations to characterize robustness.
	\end{enumerate}
	
	\section{Framework Synthesis and Methodological Implications}
	
	The proposed framework provides a domain-neutral architecture for adaptive
	multi-agent learning systems. By integrating synthetic populations,
	structured environments, survey-informed behavioral priors, causal graphs,
	information-theoretic diagnostics, and unsupervised clustering, it extends
	the interpretive and diagnostic capabilities of MAS beyond static scenario
	analysis.
	
	The four-regime structure (CPCA, CPVA, VPCA, VPVA) clarifies where
	methodological gaps in the MAS and ABM literature lie: while CPCA and CPVA
	correspond to standard forward simulations with fixed controls, VPCA and
	VPVA address the less studied case where both agents and system-level
	parameters adapt. This is precisely where decision drift, unintended
	consequences, and complex feedbacks are most likely to arise, and where
	formal diagnostics and causal explanations are most needed.
	
	From an AI perspective, the framework can be seen as a unifying template
	for combining multi-agent learning, external control optimization,
	information-theoretic evaluation, and causal reasoning. It does not
	prescribe a specific learning algorithm or optimizer, but rather specifies
	how such components can be composed and analyzed within a single
	architecture.
	
	Because the framework is deliberately domain-neutral, it can be
	instantiated in multiple application areas without changing the
	methodological core. Concrete instantiations would require specifying
	performance functionals, data sources for IPF, survey instruments, and
	interaction graphs, but the layered structure, regime taxonomy, and
	diagnostic toolkit remain the same.
	
	\section{Case Study: Emissions Policy as Adaptive Load Balancing}
\label{sec:emissions}

To illustrate how the proposed framework applies to a general policy
problem, we consider emissions regulation as a load-balancing system.
Emissions constitute a shared, capacity-limited resource: economic agents
generate emissions through production or consumption, while a policymaker
sets a cap, tax, or subsidy structure to maintain environmental
sustainability. The resulting dynamics exhibit feedback, adaptation,
bounded rationality, and long-run path dependence, making emissions
policy a natural instantiation of the four-regime architecture.

\subsection{Model Definition}

We consider a population of $N$ agents generating emissions over discrete
time steps $t = 1,\dots,T$. Let $e_{i,t}$ denote the emissions of agent $i$
at time $t$. Each agent has attributes $(\theta_i,\eta_i)$ describing
technological efficiency $\theta_i$ and propensity to adopt cleaner
alternatives $\eta_i$.

Agents choose an emissions-generating action
\[
x_{i,t} = f(\theta_i, \eta_i, P_t, s_t),
\]
where $P_t$ is a vector of policy parameters (e.g., carbon tax, cap,
subsidy) and $s_t$ is the system state, which may include past emissions or
enforcement signals. Emissions resulting from the action satisfy
\[
e_{i,t} = g(x_{i,t},\theta_i).
\]

Aggregate emissions at time $t$ are
\[
E_t = \sum_{i=1}^N e_{i,t},
\]
subject to a system-level capacity constraint
\[
E_t \le C_t,
\]
where $C_t$ is an emissions cap or adaptive environmental budget.

The system state is $s_t = E_t$ (or a richer vector including volatility
or compliance indicators). Performance balances sustainability, economic
cost, and stability via a scalar functional
\[
\Phi(s_t)
= -\alpha E_t - \beta O_t - \gamma V_t,
\]
where $O_t$ measures the frequency or severity of cap exceedances and $V_t$
captures volatility in emissions or compliance.

Over a finite evaluation window of length $K$, the overall performance of
policy parameters $P$ under learning dynamics $L$ is
\[
J(P;L) = \frac{1}{K}
\sum_{t=T-K+1}^{T} \Phi(s_t).
\]

\subsection{Population Layer: Synthetic Emitters}

A synthetic population of firms or households is generated via IPF from
aggregate environmental accounts, sectoral inventories, or survey data.
Attributes $(\theta_i, \eta_i)$ encode heterogeneity in technology,
abatement potential, and behavioral responsiveness. This representation is
domain-neutral: the population may represent industries, transport modes,
or households without loss of generality.

\subsection{Environment Layer: Emissions Capacity and Sectors}

The environment is defined by an emissions capacity $C_t$ and optionally a
sectoral structure. Let $S = \{1,\dots,M\}$ denote sectors. Each sector
$j$ has a capacity $C_{j,t}$ and receives emissions from agents
$N(j)$. Aggregate emissions satisfy:
\[
E_{j,t} = \sum_{i \in N(j)} e_{i,t},
\qquad
E_t = \sum_{j=1}^M E_{j,t}.
\]
This parallels the load on nodes in a distribution network, but without
spatial geometry.

\subsection{Behavioral Layer: Adaptive Abatement Decisions}

In the static case, emissions follow baseline technological efficiency:
\[
e_{i,t} = g(\theta_i).
\]
In the adaptive case, agents adjust emissions in response to policy signals:
\[
e_{i,t+1}
= e_{i,t} - \eta_i \bigl( P_t + c_t \bigr),
\]
where $c_t$ is a congestion signal derived from proximity to the cap
(e.g., marginal damage cost or a scarcity surcharge when $E_t$ nears
$C_t$). This formulation captures bounded rationality, reinforcement
learning, or threshold-based adoption of cleaner technologies.

\subsection{Policy Layer: Adaptive Regulation}

Policy parameters are represented as:
\[
P_t = (\lambda_t, \tau_t, \sigma_t),
\]
where $\lambda_t$ is a carbon tax or price, $\tau_t$ a cap or emissions
budget, and $\sigma_t$ a subsidy or support parameter. The policymaker
updates $P_t$ via an optimization rule $G$ using observed performance:
\[
P_{t+1} = G(P_t, \hat{J}_t, s_t).
\]
This captures iterative adjustments common in climate policy, such as
updating carbon prices or tightening emissions caps.

\subsection{Regimes Instantiation}

The emissions framework instantiates the four regimes as follows:

\paragraph{CPCA: Constant Policy, Constant Agents.}
$\lambda_t, \tau_t, \sigma_t$ fixed; no technological learning ($\eta_i = 0$).
Represents baseline or static compliance scenarios.

\paragraph{CPVA: Constant Policy, Variable Agents.}
Policy fixed; agents adapt via efficiency gains or technology adoption
($\eta_i > 0$).

\paragraph{VPCA: Variable Policy, Constant Agents.}
Policymaker adapts $P_t$; firms do not change technology.

\paragraph{VPVA: Variable Policy, Variable Agents.}
Both policy and behavior adapt; fundamental feedbacks emerge, often yielding
oscillatory or drifting emissions trajectories.

\subsection{SCM Representation}

We define an SCM with variables:
\begin{itemize}
	\item $X_t$: exogenous drivers (economic activity, shocks),
	\item $\Theta_i$: agent attributes $(\theta_i, \eta_i)$,
	\item $P_t$: policy parameters,
	\item $E_t$: aggregate emissions,
	\item $Y_t$: welfare outcomes (cost, compliance, volatility).
\end{itemize}
Directed edges include $(P_t, \Theta, X_t) \to E_t$ and $E_t \to Y_t$, while
policy adaptation introduces $E_t \to P_{t+1}$. Interventions
$\mathrm{do}(P_t = p)$ formalize counterfactuals about alternate tax or cap
trajectories.

\subsection{Diagnostics: Information and Structure}

Although the raw data originate from agent-level emissions paths 
$\{e_{i,t}\}$, the information-theoretic diagnostics are computed from 
aggregate observables derived from these micro-level actions. Individual 
emissions are first aggregated to produce a system-wide emissions time 
series $E_t = \sum_i e_{i,t}$, or analogous sectoral aggregates, and it 
is these trajectories that are used to estimate $h_\mu$, $C_\mu$, and $E$. 
Once computed, these quantities become \emph{run-level} descriptors of the 
dynamical behavior of the system rather than agent-level metrics. They 
stand alongside macro-indicators such as mean emissions, overload 
frequency, proximity to the cap, and volatility, forming a unified set of 
summary statistics for each simulation configuration. A full methodological treatment of $\epsilon$-machine reconstruction and complexity profiling in MAS is presented in \citep{garrone2025}. This makes it possible to cluster \emph{complete simulation runs} to identify distinct 
dynamic patterns and to classify emergent emission-regime types.

This run-level clustering complements traditional agent-level clustering 
that is often used in ABM to identify behavioral or socio-demographic 
agent types. While micro-level clustering groups agents according to 
traits, propensities, or their time-averaged emissions behavior, macro-level 
clustering groups simulation outcomes into dynamic regimes (stable, 
near-critical, oscillatory, or unstable). When used together, the two 
approaches allow researchers to link heterogeneity in agent roles—such as 
high emitters, responsive adopters, or inertia-prone agents—to the macro 
regimes identified across runs. This establishes a bridge between 
population composition and the emergent structure of system-wide 
dynamics.

Time series of $E_t$ or sectoral emissions are analyzed through:
\begin{itemize}
	\item entropy rate $h_\mu$ for unpredictability,
	\item statistical complexity $C_\mu$ for structural richness,
	\item predictive information $E$ for regime transitions.
\end{itemize}

Clustering of run-level summary statistics (e.g., mean 
emissions, overload frequency, $h_\mu$, $C_\mu$, $E$) identifies stable, 
near-critical, and unstable emission regimes, revealing how combinations 
of learning behavior and policy search shape the resulting trajectory 
classes. Near-cap operation induces increases in $h_\mu$ and $C_\mu$, reflecting a 
transition from stable emissions trajectories to volatile or near-chaotic 
dynamics. As agents react to tightening constraints and shifting policy 
signals, the emissions process $E_t$ becomes less predictable (higher 
$h_\mu$), more structurally rich (higher $C_\mu$), and exhibits stronger 
dependence between past and future (increasing $E$). To characterize these 
shifts, clustering (PCA + $k$-means or Gaussian mixtures) is applied to 
feature vectors combining:
\[
(\text{mean emissions},\ \text{cap exceedance frequency},\ h_\mu,\ C_\mu,\ E).
\]

The resulting clusters distinguish qualitatively different system regimes:
\begin{itemize}
	\item \textbf{stable regimes} (low emissions, low volatility),
	\item \textbf{near-critical regimes} (high $C_\mu$, emerging structural complexity),
	\item \textbf{cap-constrained or overloaded regimes} (high $h_\mu$, low predictability),
	\item \textbf{oscillatory regimes} (intermediate entropy, alternating periods of 
	abatement and rebound).
\end{itemize}

Together, these diagnostics reveal how combinations of boundedly rational 
learning behavior and adaptive policy search shape the trajectory classes 
that emerge near critical operating conditions.

\subsection{Experimental Protocol}

Experiments vary:
\begin{itemize}
	\item initial policy $(\lambda_0,\tau_0,\sigma_0)$,
	\item learning responsiveness $\eta_i$,
	\item exogenous shocks $X_t$,
	\item capacity constraints $C_t$ or sectoral budgets.
\end{itemize}
For each configuration, $R$ replications of length $T$ are run; performance
$J(P;L)$ is computed over a window $K$. Sensitivity analysis quantifies how
policy parameters and learning rates affect stability and long-run emissions.

This case demonstrates how emissions policy fits naturally into the proposed
framework as a load-balancing problem with adaptive agents, adaptive
policy, causal interpretability, and information-theoretic diagnostics.

	\section{Case Study: Adaptive Load Balancing in Electric Grids via Demand Response}
	\label{sec:smartgrid}
	
Modern electric grids increasingly
	rely on distributed control, demand response, and adaptive pricing to maintain
	stability under fluctuating loads. The resulting dynamics are well represented
	as a multi-agent system: households and firms behave as adaptive loads, while
	a system operator adjusts tariffs or control signals to prevent overload of
	transformers or feeders.
	
	\subsection{Model Definition}
	
	We consider a distribution grid with $M$ nodes (transformers or feeders),
	each with capacity $C_j$. At discrete time steps $t=1,\dots,T$, a population
	of consumers (agents) generates electricity demand. Let $a_{i,t}$ be the
	demand of agent $i$ at time $t$. Each agent has attributes $(\theta_i, \eta_i)$
	encoding baseline consumption $\theta_i$ and price responsiveness $\eta_i$.
	
	Consumers choose a time-varying consumption level
	\[
	x_{i,t} = f(\theta_i, \eta_i, P_t, c_t),
	\]
	where $P_t$ is a vector of system-level control parameters (e.g., dynamic
	tariffs) and $c_t$ is a local congestion signal depending on the load at
	the agent’s node. Consumption aggregates to node-level load:
	\[
	L_{j,t} = \sum_{i \in N(j)} x_{i,t},
	\]
	where $N(j)$ is the set of agents connected to node $j$. If $L_{j,t} > C_j$,
	the node is overloaded, causing losses or voltage drops.
	
	The system state is $s_t = (L_{1,t},\dots,L_{M,t})$. Performance balances
	stability, efficiency, and fairness using a scalar functional $\Phi(s_t)$.
	A typical choice is
	\[
	\Phi(s_t) = -\alpha D_t - \beta O_t - \gamma V_t,
	\]
	where $D_t$ is aggregate demand, $O_t$ the fraction of overloaded nodes, and
	$V_t$ a measure of voltage deviation. Over a window of size $K$,
	\[
	J(P;L) = \frac{1}{K} \sum_{t=T-K+1}^T \Phi(s_t).
	\]
	
	\subsection{Population Layer: Synthetic Consumers}
	
	A synthetic population is generated by IPF using aggregate statistics such as
	household size, appliance ownership, income class, or time-of-use patterns.
	Attributes $(\theta_i, \eta_i)$ are drawn from this population: heterogeneous
	baseline loads $\theta_i$ represent housing, climate, and lifestyle
	differences, while price responsiveness $\eta_i$ captures consumer willingness
	to shift or reduce consumption under dynamic tariffs.
	
	This layer defines heterogeneity without committing to any specific empirical
	context.
	
	\subsection{Environment Layer: Distribution Grid Topology}
	
	The environment is a graph $G=(V,E)$ where $V$ are transformers/feeders and
	$E$ represent distribution lines. Each node $j$ has capacity $C_j$ and a set
	of connected consumers. Power flows are represented in simplified form through
	node-level loads $L_{j,t}$; full AC power flow equations are not needed for
	load-balanced demand response studies.
	
	\subsection{Behavioral Layer: Consumer Adaptation}
	
	In the static case, consumption follows a fixed function $x_{i,t}=f(\theta_i)$.
	In the adaptive case, agents respond to time-varying tariffs and congestion:
	\[
	x_{i,t+1} =
	x_{i,t} - \eta_i \bigl( P_t + c_{j(i),t} \bigr)
	\]
	where $c_{j(i),t}$ is a congestion penalty at the node where agent $i$ is
	connected. This captures boundedly rational adaptation, discrete choice, or
	reinforcement learning behavior.
	
	\subsection{Policy Layer: Dynamic Tariffs and Control}
	
	The system operator adjusts tariffs $P_t$ to reduce overload. We consider two
	controls:
	\begin{enumerate}
		\item time-varying price multiplier $\lambda_t$, and
		\item congestion threshold $\tau_t$ indicating when surcharge applies.
	\end{enumerate}
	
	The control vector $P_t = (\lambda_t, \tau_t)$ is updated by a policy search
	algorithm $G$ that aims to improve $J(P;L)$. A hill-climbing or evolutionary
	strategy can serve as $G$, treating the MAS as a noisy black-box mapping.
	
	\subsection{Regime Instantiation}
	
	The four regimes are instantiated as follows.
	
	\paragraph{CPCA: Constant Control, Constant Agents.}
	$\lambda_t=\lambda$, $\tau_t=\tau$ fixed; no consumer adaptation ($\eta_i=0$).
	
	\paragraph{CPVA: Constant Control, Variable Agents.}
	Prices constant; consumers adapt ($\eta_i>0$).
	
	\paragraph{VPCA: Variable Control, Constant Agents.}
	Consumers do not adapt; the system operator searches over $P_t$.
	
	\paragraph{VPVA: Variable Control, Variable Agents.}
	Both consumers and the system operator adapt. This regime exhibits the most
	complex dynamics, including oscillations between under- and over-reaction.
	
	\subsection{SCM Representation}
	
	An SCM captures the causal structure:
	\begin{itemize}
		\item $X_t$: exogenous factors (weather, baseline demand);
		\item $P_t$: tariffs and congestion thresholds;
		\item $\Theta$: consumer attributes $(\theta_i, \eta_i)$;
		\item $s_t$: node loads and congestion;
		\item $Y_t$: performance outcomes (overload, demand, voltage).
	\end{itemize}
	
	Arrows represent relationships such as
	$(P_t, \Theta, X_t) \to s_t$ and $s_t \to Y_t$, while adaptive control adds
	$s_t \to P_{t+1}$. Interventions $\mathrm{do}(P_t=p)$ capture counterfactual
	comparisons between adaptive and static frameworks.
	
\subsection{Diagnostics: Information and Structure}

As in the previous instance, the information-theoretic diagnostics are computed from 
aggregate observables derived from node-level loads. Individual 
consumption is first aggregated to produce load trajectories $L_{j,t}$ 
over nodes $j$, and a representative system-level observable (e.g., 
total demand $D_t$ or a symbolized overload indicator) is extracted. 
It is this aggregate time series that is used to estimate $h_\mu$, 
$C_\mu$, and $E$, which then serve as \emph{run-level} summaries of the 
dynamical behavior of each simulation rather than agent-level metrics. 
Once computed, these diagnostics stand alongside macro indicators such as 
overload frequency and mean demand, enabling clustering of complete 
simulation runs to reveal distinct operational regimes. See \citep{garrone2025} for a general formulation of $\epsilon$-machine–based diagnostics in adaptive multi-agent systems.

Clustering at this run-level resolves classes of emergent system 
trajectories—for example, stable, near-critical, oscillatory, or 
overloaded regimes. This represents one natural use of clustering in 
adaptive MAS. A complementary use, common in agent-based modeling, 
clusters \emph{agents} themselves based on traits, behavioral propensities, 
or time-averaged actions. Such micro-level clustering can be used to link 
heterogeneous agent roles (e.g., high-demand households, flexible users, 
price-sensitive adopters) to the macro-level clusters identified at the 
run level. Together, macro- and micro-level clustering provide a unified 
view of how population heterogeneity shapes, and is shaped by, emergent 
system dynamics.

From the trajectories of $s_t$, we compute:
\begin{itemize}
	\item entropy rate $h_\mu$ of load dynamics,
	\item statistical complexity $C_\mu$ of the reconstructed $\epsilon$-machine,
	\item predictive information $E$ between past and future loads.
\end{itemize}

We expect information-theoretic quantities to spike when the grid operates
near capacity, reflecting a phase transition from stable to overloaded 
behavior. Clustering (PCA + $k$-means or Gaussian mixtures) is applied to 
feature vectors combining:
\[
(\text{mean demand},\ \text{overload frequency},\ h_\mu,\ C_\mu,\ E).
\]

Clusters naturally separate into:
\begin{itemize}
	\item stable regimes (low overload),
	\item near-critical regimes (high $C_\mu$),
	\item overloaded regimes (high $h_\mu$, low predictability),
	\item oscillatory regimes (intermediate entropy, cyclic patterns).
\end{itemize}

	\subsection{Experimental Protocol}
	
	A typical experiment varies:
	\begin{itemize}
		\item exogenous demand patterns (peak/off-peak),
		\item tariff initialization $(\lambda_0, \tau_0)$,
		\item agent responsiveness $\eta_i$,
		\item grid capacity constraints.
	\end{itemize}
	
	For each configuration and regime, $R$ replications of length $T$ are run, and
	$J(P;L)$ is evaluated over the last $K$ steps. Sensitivity analysis identifies
	dominant interactions between learning rates, capacities, and optimization
	parameters.
	
	This case study illustrates how the proposed framework integrates adaptive
	behavior, system-level control, causal interpretation, and information-
	theoretic diagnostics in a realistic policy-relevant setting.
	
\section{Synthesis and Discussion}
The two case studies illustrate the generality and transferability of the proposed framework across both policy and infrastructure domains. Their high-level motivations, summarized in Table~\ref{tab:purpose}, show that despite addressing substantively different contexts—environmental emissions regulation and electric-grid demand management—each case instantiates the same core problem structure: resource constraints, adaptive agents, and an adaptive controller. Their structural parallelism is intentional. 

\begin{table}[ht]
	\centering
	\small
	\begin{tabular}{p{2cm} p{6.6cm} p{6.6cm}}
		\toprule
		\textbf{Aspect} & \textbf{Emissions Policy Case} & \textbf{Electricity Load-Balancing Case} \\
		\midrule
		Motivation &
		Environmental regulation; managing pollution within sustainable limits. &
		Grid reliability; avoiding overloads and managing peak demand. \\
		
		Resource Mapped &
		Emissions treated as a load on a capacity-limited environmental system. &
		Electric load mapped to transformer/feeder capacity constraints. \\
		
		Agents &
		Firms or households generating emissions; respond to policy incentives. &
		Households and firms generating electricity demand; respond to tariffs. \\
		
		Regulator &
		Environmental authority adjusting taxes, caps, subsidies. &
		System operator (DSO) adjusting dynamic tariffs, thresholds. \\
		
		Primary Goal &
		Maintain sustainability and prevent exceeding environmental capacity. &
		Maintain grid stability and avoid transformer/feeder overloads. \\
		\bottomrule
	\end{tabular}
	\caption{High-level motivation of the two case studies. Both instantiate the same conceptual machinery—synthetic populations, boundedly rational agents, adaptive control, policy search, and diagnostic tools—demonstrating methodological generality across policy and infrastructure domains.}
	\label{tab:purpose}
\end{table}

In the emissions case, pollution output plays the role of a load on a shared environmental capacity, while abatement decisions correspond to reductions in that load; in the electricity case, household consumption contributes to nodal loads, and demand shifting plays an analogous role to abatement. Likewise, taxes, caps, and subsidies mirror dynamic tariffs and congestion thresholds, and the environmental regulator parallels the grid operator. This isomorphism demonstrates that the framework abstracts from domain-specific semantics, enabling a uniform treatment of adaptation, control, and emergent behavior.

\begin{table}[ht]
	\centering
	\small
	\begin{tabular}{p{2cm} p{6.6cm} p{6.6cm}}
		\toprule
		\textbf{Aspect} & \textbf{Emissions Policy Case} & \textbf{Electricity Load-Balancing Case} \\
		\midrule
		Agent Attributes &
		Technological efficiency $\theta_i$; responsiveness to clean alternatives $\eta_i$. &
		Baseline load $\theta_i$; price responsiveness $\eta_i$. \\
		
		Adaptive Behavior &
		Agents reduce emissions based on taxes, congestion (proximity to cap), and responsiveness. &
		Agents reduce or shift consumption based on dynamic tariffs and local congestion. \\
		
		Update Rule &
		$e_{i,t+1} = e_{i,t} - \eta_i(P_t + \text{congestion})$ &
		$x_{i,t+1} = x_{i,t} - \eta_i(P_t + \text{congestion})$ \\
		
		Functional Interpretation &
		Abatement effort; cleaner technology adoption; behavioral adjustment. &
		Demand shifting; peak shaving; response to real-time price signals. \\
		\bottomrule
	\end{tabular}
	\caption{Comparison of agent attributes and behavioral updates. Both domains use parallel adaptive rules, differing only in interpretation: emissions abatement versus electricity demand shifting.}
	\label{tab:agents}
\end{table}

At the methodological level, both cases rely on the same diagnostic stack—structural causal models for intervention semantics, information-theoretic measures for detecting shifts in predictability and latent structure, and clustering techniques for identifying emergent dynamic regimes. The parallel structure of agent attributes and adaptive behavior (Table~\ref{tab:agents}) underscores how the same behavioral update equation is instantiated in two semantically distinct domains. Differences in environmental representation (Table~\ref{tab:environment}) highlight the shift from an abstract, sector-based capacity constraint to a fully spatial, networked topology with node-specific limits. Likewise, distinctions in the policy and control layers (Table~\ref{tab:policy}) show how regulatory instruments and operational tariffs can be expressed within a unified control vector and adapted through the same optimization mechanism.

\begin{table}[ht]
	\centering
	\small
	\begin{tabular}{p{2cm} p{6.6cm} p{6.6cm}}
		\toprule
		\textbf{Feature} & \textbf{Emissions Policy Case} & \textbf{Electricity Load-Balancing Case} \\
		\midrule
		Topology &
		Non-spatial; sectoral or aggregate population. &
		Explicit network graph $G=(V,E)$; node-specific agents. \\
		
		Capacity Structure &
		Global or sector-specific capacity $C_t$. &
		Node-level capacities (transformers/feeders) $C_j$. \\
		
		Congestion Mechanism &
		Exceeding or approaching emissions cap triggers policy pressure. &
		Local overload occurs when demand $>$ node capacity $C_j$. \\
		
		Spatiality &
		Abstract; no geometry required. &
		Strong spatial component; topology shapes agent interactions. \\
		\bottomrule
	\end{tabular}
	\caption{Comparison of environmental structures. The emissions model uses an abstract capacity constraint, whereas the electricity model embeds agents in a physical network, adding spatial heterogeneity and localized congestion.}
	\label{tab:environment}
\end{table}

\begin{table}[ht]
	\centering
	\small
	\begin{tabular}{p{2cm} p{6.6cm} p{6.6cm}}
		\toprule
		\textbf{Aspect} & \textbf{Emissions Policy Case} & \textbf{Electricity Load-Balancing Case} \\
		\midrule
		Policy Vector Components &
		$(\lambda_t, \tau_t, \sigma_t)$: carbon tax, emissions cap, subsidy. &
		$(\lambda_t, \tau_t)$: price multiplier, congestion threshold. \\
		
		Control Objective &
		Regulate emissions intensity and compliance with environmental limits. &
		Maintain grid stability and reduce peak load. \\
		
		Feedback Loop &
		Policy reacts to aggregate emissions and volatility. &
		Operator reacts to nodal overload and grid stress. \\
		
		Dimensionality &
		More multi-dimensional (three levers). &
		More operational (tariff + threshold). \\
		
		Adaptive Search &
		External search adjusts policy vector to improve performance metrics. &
		Identical adaptive search structure applied to grid-control parameters. \\
		\bottomrule
	\end{tabular}
	\caption{Comparison of policy/control layers. Both treat policy as a dynamic control variable adapted via external optimization, but the emissions domain centers on regulatory instruments while the electricity case focuses on operational grid management.}
	\label{tab:policy}
\end{table}

Across both domains, the diagnostic tools reveal consistent signatures of stability, criticality, and oscillatory behavior. Because both case studies can be run under the CPCA, CPVA, VPCA, and VPVA regimes, they provide a comparative view of how combinations of agent adaptation and policy search shape system dynamics. The emissions case highlights policy drift, long-run sustainability constraints, and macro-level volatility, whereas the electricity case emphasizes operational stability, network congestion, and real-time adaptation. Together, these contrasts reinforce the claim that the framework is domain-neutral and provides a general methodological lens for analyzing adaptive multi-agent systems under dynamic policy and resource constraints.

\section{Conclusion}

This paper has presented a general framework for adaptive multi-agent
learning in systems where both agents and policy-makers co-evolve over
time. The approach combines four key components: (i) a taxonomy of dynamic
regimes describing the joint adaptation of agents and system-level control
parameters; (ii) the integration of synthetic population methods, structured
interaction topologies, and survey-informed priors as modular initialization
elements; (iii) causal and information-theoretic diagnostics for assessing
predictability, stability, and structural change in generated trajectories;
and (iv) clustering techniques for uncovering emergent regimes in
high-dimensional output spaces.

Taken together, these elements provide a domain-neutral blueprint for
constructing, analyzing, and explaining adaptive multi-agent systems. By
separating agent behavior, learning rules, system-wide policy adaptation,
and diagnostic tools into modular components, the framework enables
systematic exploration of how local decision rules and adaptive control
interact to produce global patterns. The design is intentionally
transparent: each component—initialization, adaptation, control, and
evaluation—can be independently modified or extended, supporting a wide
range of methodological and applied research.

Future work will apply the framework to concrete MAS settings to evaluate
its performance relative to static or single-regime designs, examine its
robustness under richer behavioral heterogeneity, and explore the benefits
of multi-level or hierarchical control architectures. Beyond methodological
advances, the framework aims to contribute practical tools for constructing
explainable and contestable decision processes in complex environments
involving adaptation, uncertainty, and policy feedback.

\section*{Acknowledgments and Disclosure Statements}

\subsection*{Preprint Statement}
This manuscript is released as a preprint to facilitate open scientific discussion and to provide a transparent methodological foundation for ongoing work. The content may undergo further revision in subsequent journal submissions.

\subsection*{PhD Research Disclosure}
This work was conducted as part of the author's doctoral research in the Faculty of Pure and Applied Sciences at the Open University of Cyprus and in affiliation with the Department of Informatics, Systems and Communication (DISCo) at the University of Milano--Bicocca. The methodological framework presented here forms part of the doctoral research agenda on adaptive multi-agent systems, causal diagnostics, and explainable policy design.

\subsection*{Acknowledgments}
The author is grateful to \textbf{Dr.\ Loizos Michael} for insightful discussions during the early stages of this work, particularly regarding the articulation of adaptive regimes and the role of formal diagnostics in multi-agent systems. His feedback helped refine the conceptual emphasis of the framework; all remaining errors or omissions are the author's responsibility.

\subsection*{Funding and Institutional Support}
No dedicated funding was received specifically for preparing this manuscript. Institutional support was provided through standard doctoral research resources from the Open University of Cyprus and the University of Milano--Bicocca. No external grants, contracts, or third-party funding mechanisms influenced the design, execution, or reporting of this work.

\subsection*{Author Contributions}
The author is solely responsible for the conception, design, mathematical formulation, implementation, analysis, and writing of this work, including all revisions.

\subsection*{Use of Digital Assistants}
During manuscript preparation, the author used digital assistants (ChatGPT, OpenAI; Perplexity.ai, San Francisco, CA; and Publish or Perish, Harzing.com) solely for language refinement, \LaTeX{} and code editing, literature retrieval, and bibliometric verification. The author reviewed and edited all generated content and assumes full responsibility for the final text and its scientific interpretations. No financial or material support, administrative assistance, or in-kind contributions were received beyond the software tools explicitly mentioned above.

\subsection*{Conflict of Interest Statement}
The author declares no conflicts of interest, financial or otherwise, related to the subject matter of this manuscript.

\subsection*{Data and Code Availability}
All code, model specifications, and computational procedures referenced in this manuscript will be made available in a public repository upon reasonable request. No proprietary or sensitive datasets were used in this study.

\end{document}